\documentclass[aps,prl,showpacs,reprint,amsmath,amssymb,groupaddresses,superscriptaddress]{revtex4-1}

\usepackage{graphicx}
\usepackage{dcolumn}
\usepackage{bm}

\begin{document}


\title{Simultaneous Measurement of Resistively and Optically Detected Nuclear Magnetic Resonance 
in the $\nu=2/3$ Fractional Quantum Hall Regime}

\author{Keiichirou Akiba}
\affiliation{Department of Physics, Tohoku University, Sendai, 980-8578, Japan}
\affiliation{JST, ERATO Nuclear Spin Electronics Project, Sendai, 980-8578, Japan}
\affiliation{Department of Applied Physics, Tokyo University of Agriculture and Technology, Koganei, Tokyo, 184-8588, Japan}

\author{Katsumi Nagase}
\affiliation{Department of Physics, Tohoku University, Sendai, 980-8578, Japan}
\affiliation{JST, ERATO Nuclear Spin Electronics Project, Sendai, 980-8578, Japan}

\author{Yoshiro Hirayama}
\affiliation{Department of Physics, Tohoku University, Sendai, 980-8578, Japan}
\affiliation{JST, ERATO Nuclear Spin Electronics Project, Sendai, 980-8578, Japan}
\affiliation{WPI-AIMR, Tohoku University, Sendai, 980-0812, Japan}


\date{\today}

\begin{abstract}
We observe nuclear magnetic resonance (NMR)  
in the fractional quantum Hall regime at Landau level filling factor $\nu=2/3$
from simultaneous measurement of longitudinal resistance and photoluminescence (PL). 
The dynamic nuclear spin polarization is induced  
by applying a huge electronic current at the  spin phase transition point of $\nu=2/3$. 
The NMR spectra obtained from changes in resistance and PL intensity 
are qualitatively the same; that is,  
the Knight shift (spin polarized region) 
and zero-shift (spin unpolarized region) resonances are observed in both.  
The observed change in PL intensity is interpreted 
as a consequence of the trion scattering induced by polarized nuclear spins. 
We conclude that both detection methods probe almost the same local phenomena.

\end{abstract}

\pacs{76.60.-k; 73.43.-f; 71.35.Pq}


\maketitle

Quantum Hall effect has attracted a lot of physical interest since its discovery, 
and it has been investigated by using various kinds of experimental methods 
as well as other phenomena of condensed matter physics. 
Different experimental techniques can usually offer a good understanding of physics. 
However, considerable discrepancies between them occasionally arise in the quantum Hall system, 
even when sample preparations and experimental conditions are almost the same. 
For instance, the different size of skyrmion has been observed and argued \cite{SkyrmionBarrett, SkyrmionAifer, SkyrmionRD, SkyrmionOPKukushkin}. 
The optical nuclear polarization observed in optical method is much larger than that in conventional and resistive methods \cite{SkyrmionOPKukushkin, OPBarrett, OPDavies, OPFukuoka, OPAkiba, OPPRB}. 
The case of the electron spin polarization at Landau level filling factor $\nu=5/2$ is more complicated; 
the different experimental methods show 
the fully polarized state, unpolarized system, and partially polarized domains 
\cite{5/2Sarma, 5/2Stern1, 5/2Rhone, 5/2NMR, 5/2Stern2, 5/2Wurstbauer, 5/2Friess, 5/2Review}. 
In these studies, the results obtained with optical methods 
especially exhibited controversial disagreements with other experimental methods.
The possible explanation of such disagreements can be expected as follows: 
optically accessible phenomena can occur in the spatially limited region and/or 
photoexcited holes can considerably affect the system due to the Coulomb interaction. 

To investigate such different experimentally-observed results, 
the simultaneous measurement by different experimental methods is 
effective. 
In this paper, we measure the resistance and photoluminescence (PL) 
in the $\nu=2/3$ fractional quantum Hall regime simultaneously. 
At $\nu=2/3$, an electron spin phase transition (SPT) can occur 
due to competition between Coulomb and Zeeman energies \cite{SpinPhysics}, 
and this SPT has been observed from 
the resistance and PL so far \cite{SpinPhysics, SPKukushkin, Hayakawa}. 
Associated with the two electron spin phases (i.e., spin polarized and unpolarized states), 
nuclear spins are polarized when a huge electronic current is applied \cite{SpinPhysics}. 
In the present study, the target phenomena to be simultaneously measured  
are this current-induced nuclear spin polarization and its nuclear magnetic resonance (NMR). 
This is because NMR provides local information from its spectrum \cite{Notelocal}, 
which has a possibility to identify the essential difference 
between resistive and optical detections. 
In addition, the still-ambiguous details of the nuclear spin polarization at $\nu=2/3$ SPT, 
which are crucial for a future application to quantum information technology \cite{YusaNature},   
can be investigated. 

We demonstrate simultaneous measurement of resistively and optically detected NMR 
with the current-induced nuclear spin polarization at $\nu=2/3$ SPT. 
Accompanying the dynamic nuclear spin polarization, 
the previously-reported enhancement of longitudinal resistance \cite{SpinPhysics} 
and the variation of PL intensity occurred at the same time. 
Subsequently, we obtained NMR spectra from changes in both resistance and PL intensity. 
The simultaneously-measured spectra are qualitatively the same. 
The resistively-detected spectrum is consistent with that of a previous study \cite{2/3RDNMR}, 
excluding the influence of optical illumination. 
The optically-detected spectra enable us to interpret that 
the variation of PL intensity due to nuclear spin polarization 
is caused by the trion (photoexcited particle) scattering. 
It is thus concluded that 
the proposed simultaneous measurement (namely, resistive and optical detection methods) probe almost the same local phenomenon.

Experiments were carried out on a single 18-nm GaAs/Al$_{0.33}$Ga$_{0.67}$As quantum well, 
which was processed to a 100-$\mu$m long and 30-$\mu$m wide Hall bar. 
The electron density $n_s$ can be controlled 
by applying a voltage to a $n$-type GaAs substrate (back gate).
The electron mobility is 185~m$^2$/(Vs) for $n_s=1.2\times 10^{15}$~m$^{-2}$. 
This sample was cooled down to 0.3~K in a cryogen free $^3$He refrigerator. 
A longitudinal resistance was measured 
by using a lock-in technique with a low-frequency (83~Hz) constant current. 
Luminescence was excited by a linearly-polarized output of 
a mode-locked Ti:sapphire laser (pulse width: $\sim 2$ ps, pulse repetition: 76 MHz) 
with wavelength of 784~nm and average power density of 2~mW/cm$^2$. 
A laser beam (diameter: $\sim$230~$\mu$m) continuously irradiated the whole Hall bar 
through an optical window on the bottom of the cryostat. 
The propagation direction of the laser beam was parallel to an external magnetic field of 7.15~T, 
which was perpendicular to the quantum well. 
The left circularly polarized ($\sigma^-$) PL was collected 
from the entire laser-excitation area through the same optical window, 
where the PL collection time was 265~s.  
The details of this experimental setup are the same as those stated in our previous work \cite{Detail}. 

\begin{figure}
\includegraphics[width=1\columnwidth]{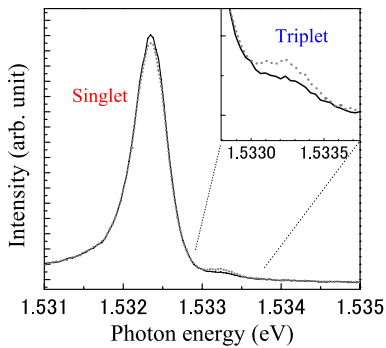}%
\caption{(Color online) Photoluminescence spectra at $\nu=2/3$ 
before (solid line) and after (dotted line) current-induced nuclear spin polarization. 
\label{fig:PL.eps}}
\end{figure}

The solid line in Fig.~\ref{fig:PL.eps} shows the PL spectrum at $\nu=2/3$. 
We observe two peaks: an apparent peak at 1.5324~eV, and a tiny peak at 1.5333~eV. 
The optical spectrum in the fractional quantum Hall regime has been understood by 
existing bound electron-hole complexes, e.g., neutral and charged (trions) excitons \cite{BarJoseph}. 
Here, the two electrons in a trion form singlet and triplet spin states.  
The lower and higher energy peaks are respectively assigned to 
singlet and triplet trion peaks \cite{Hayakawa, OPPRL}. 

\begin{figure}
\includegraphics[width=1\columnwidth]{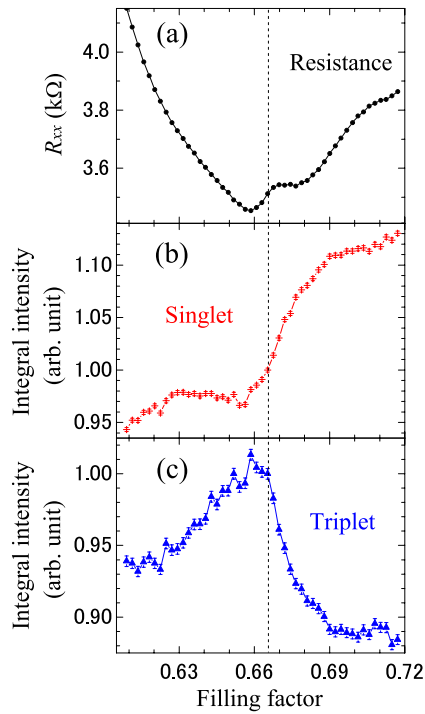}%
\caption{(Color online) Simultaneous measurement of (a) longitudinal resistance $R_{xx}$, 
(b) integral singlet intensity, and (c) integral triplet intensity around $\nu=2/3$. 
The error bars show the standard deviations of total counts.  
\label{fig:SPT.eps}}
\end{figure}

Figure \ref{fig:SPT.eps} shows the simultaneously measured  
longitudinal resistance $R_{xx}$ and PL around $\nu=2/3$.  
The applied current was 30~nA, which was low enough not to polarize nuclear spins, 
and $\nu$ was tuned by using the back gate in the fixed magnetic field. 
The large dip in $R_{xx}$ in (a) is associated with the  fractional quantum Hall state at $\nu=2/3$
and the $R_{xx}$ peak at $\nu\sim 0.67$ is caused by the spin phase transition \cite{SpinPhysics}. 
The spin polarized (unpolarized) state is known to be formed 
on the lower (higher) $\nu$ side of this peak. 
As shown in Fig.~\ref{fig:PL.eps}, the singlet and triple peaks appear in the PL spectrum. 
The integral intensity around these peaks in (b) and (c) was recorded. 
As $\nu$ increases, the singlet intensity starts to increase around the $R_{xx}$ peak. 
This increase in singlet intensity accompanies with a decrease in triplet intensity. 
These changes in PL intensity are consistent with the SPT; 
the triplet (singlet) trion mainly should reside in the spin polarized (unpolarized) region 
because the two electron spins in a triplet (singlet) trion are aligned in parallel (anti-parallel).  
Indeed, our observation of the SPT from PL is in qualitative agreement with 
a previous investigation on optical detection of the spin phase transition \cite{Hayakawa, Note1}. 

\begin{figure}
\includegraphics[width=1\columnwidth]{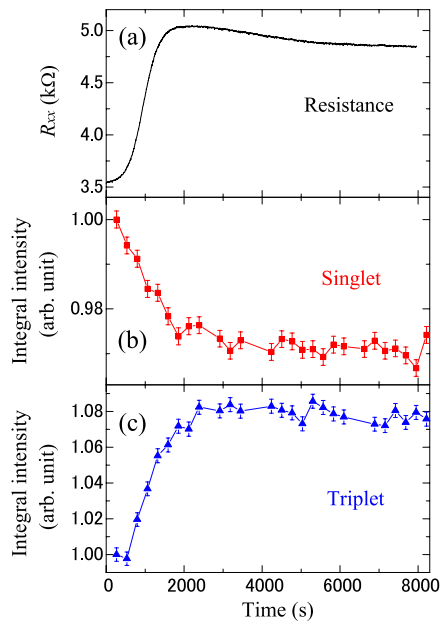}%
\caption{(Color online) Simultaneous measurement of (a) longitudinal resistance $R_{xx}$, 
(b) integral singlet intensity, and (c) integral triplet intensity with the current-induced nuclear spin polarization at $\nu=2/3$. The error bars show the standard deviations of total counts.
\label{fig:TimeEvo.eps}}
\end{figure}

In order to induce dynamic nuclear spin polarization, 
we applied a huge current of 240~nA at the $\nu=2/3$ SPT point (the broken line in Fig.~\ref{fig:SPT.eps}). 
After the current-induced nuclear spin polarization (polarization time: 8000~s), 
singlet intensity decreases and triplet intensity increases 
as shown by the dotted line in Fig.~\ref{fig:PL.eps}. 
The temporal development of these PL intensity changes is presented in Fig.~\ref{fig:TimeEvo.eps}, 
where $R_{xx}$ was measured simultaneously. 
As already reported \cite{SpinPhysics}, 
the resistance enhancement occurs due to the nuclear spin polarization. 
The time scale of the change in PL integral intensity 
(i.e., the decrease in singlet integral intensity and the increase in triplet integral intensity) 
is almost the same as that of the resistance enhancement (see Fig.~\ref{fig:TimeEvo.eps}).
This strongly suggests that 
the PL intensity changes are caused by the current-induced nuclear spin polarization. 

As mentioned above, 
the triplet (singlet) trion mainly should reside in the spin polarized (unpolarized) region. 
Therefore, 
the increase in triplet intensity accompanying the decrease in singlet intensity 
in Fig.~\ref{fig:TimeEvo.eps}
seems to mean 
both spreading of the spin polarized region and shrinking of the spin unpolarized region. 
However, the saturated value of triplet integral intensity 
is larger than the maximum value of that in Fig.~\ref{fig:SPT.eps} (c); 
that is, triplet intensity after nuclear spin polarization 
is larger than that in the spin polarized state.  
The total amount of PL intensity with dynamic nuclear spin polarization 
is attributed to not only the size of the spin region but also another factor.  
In a later section, we will discuss the origin of the change in PL intensity.

\begin{figure}
\includegraphics[width=1\columnwidth]{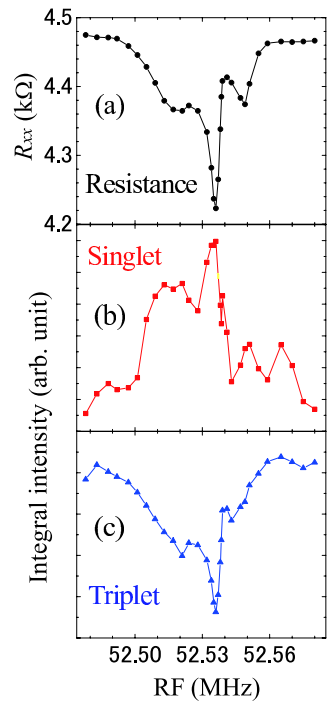}%
\caption{(Color online) NMR spectra obtained from (a) longitudinal resistance $R_{xx}$, (b) singlet integral intensity, and (c) triplet integral intensity for $^{75}$As nuclear spin. 
\label{fig:NMR.eps}}
\end{figure}

Next, we performed the NMR experiment. 
To obtain the NMR spectra, we used the following procedures. 
First, a huge current (240~nA) was applied to the sample at $\nu=2/3$  
for long enough to saturate the resistance change (over 3~h), 
where an off-resonant radio frequency (RF) magnetic field was irradiated using a handmade split coil 
in order to incorporate the influence of the RF irradiation (e.g. electron temperature increase). 
Second, we only changed the RF frequency 
and then waited for 250~s so that the system reached a stationary state. 
Third, the PL spectra were collected for 265~s and $R_{xx}$ was measured 100 times during the PL collection time. 
As a result, we acquired an averaged $R_{xx}$ 
and singlet and triplet integral intensities 
at a certain RF. 
By repeating the second and third procedures, 
we obtained the NMR spectra from the changes in resistance and PL intensity simultaneously. 

Figure~\ref{fig:NMR.eps} shows the NMR spectra for $^{75}$As nuclear spin~$I=3/2$, 
where 40 spectra obtained by each detection method were averaged, 
since the signal-to-noise ratio was too low for PL detection \cite{SN, Note2}.  
As shown in (a), the resistively-detected NMR spectrum clearly exhibits
two relatively sharp resonance lines at 52.536 and 52.549~MHz 
and one broadened resonance peak on the lower-frequency side.   
These features are also observed in the (b) singlet and (c) triplet PL detections, 
where the signal in (b) varies in the opposite direction 
because the change in singlet intensity associated with the nuclear spin polarization 
is opposite to the others (see Fig.~\ref{fig:TimeEvo.eps}). 
Although the signals obtained from the PL were influenced by the PL intensity fluctuation, 
all spectra (a)--(c) are qualitatively the same \cite{comment1}. 
Note that we demonstrate for the first time 
not only the simultaneous measurement of the resistive and optical NMR 
but also the optical NMR from trion intensity.

The sharp and broad resonances are respectively attributed to 
the spin unpolarized and polarized regions. 
The energy of nuclear spin resonance is shifted by the electron spin polarization $\mathcal{P}_e$
due to hyperfine interaction, which is known as the Knight shift \cite{SpinPhysics}. 
The spin unpolarized region ($\mathcal{P}_e=0$) brings about no energy shift
and the spin polarized region ($\mathcal{P}_e=1$) causes a negative energy shift. 
This energy shift depends on electron density; 
the electron distribution in the growth direction, 
which is formed by confinement of the quantum well, 
broadens the total Knight shift \cite{Tycko}. 
We interpret two of the sharp resonance lines 
as a consequence of quadrupole splitting and 
the population difference among fourfold-nuclear-spin levels. 
Although the quadrupole splitting of $^{75}$As nuclear spins causes three resonance lines, 
the population distribution among four levels changes the relative strengths of these lines.  
The fact that two Knight shift resonances were not observed is due to the broadening. 
The frequency scales of the broadening and the splitting 
(i.e., the total shape of the NMR specrtum) are consistent with 
the previous studies on resistively-detected NMR 
of $\nu=2/3$ \cite{2/3RDNMR, Kronmuller, Kumada, Note3}. 
Moreover, the consistency of the previous reports means that 
the laser illumination does not influence resistively-detected results 
in spite of the existence of photoexicted carriers.

On the basis of our interpretation of the NMR spectra,
we consider physics behind the optical detection as follows. 

The NMR spectra obtained 
by both the singlet [Fig.~\ref{fig:NMR.eps}(b)] and triplet [Fig.~\ref{fig:NMR.eps}(c)] PLs 
show the contributions from the spin unpolarized and polarized regions. 
This fact indicates that 
the singlet and triplet trions coexist in the spin polarized and also unpolarized regions 
after the nuclear spin polarization. 

We, here, consider the strength of the NMR signal. 
In Figs.~\ref{fig:NMR.eps} (b) and (c), 
neither the spin polarized nor unpolarized signal is striking.
This observation is not expected and might seem to be a contradiction; that is, 
although the singlet (triplet) trions reside in both the spin polarized and unpolarized regions, 
the singlet (triplet) trion preferably exists in the spin unpolarized (polarized) region 
and the spin unpolarized (polarized) signal should be pronounced in (b) [(c)].
However, we only recorded the deviations from the change in PL intensity 
accompanying the nuclear spin polarization in (b) [(c)]. 
Therefore, when the nuclear spin polarization simply 
decreases singlet intensity and simultaneously increases triplet intensity 
in both the spin polarized and unpolarized regions, 
we can understand the seemingly contradicting result
even though the spin unpolarized (polarized) region mainly radiates singlet (triplet) trion light. 
This means that the change in singlet (triplet) PL intensity due to nuclear spin polarization
does not solely depend on the size of the spin unpolarized (polarized) region.

We interpret the change in PL intensity as trion scattering 
induced by the nuclear spin polarization. 
After the current-induced nuclear spin polarization, 
the observed $R_{xx}$ peak around $\nu=2/3$ becomes larger and broader, 
indicating that 
the obtained nuclear spin polarization is spatially inhomogeneous 
and that both positive and negative polarizations exist  
\cite{SpinPhysics, Kraus}. 
Therefore, the current-induced nuclear spin polarization creates 
a spatial modulation of the electron Zeeman energy 
through the hyperfine interaction. 
This potential fluctuation can enhance the trion-scattering process. 
The scattering usually suppresses radiative recombination, 
which accounts for the change in singlet PL intensity. 
In the case of the triplet trion, the scattering should enhance PL intensity 
in order to explain the experimental results.
This is understood by the existence of the dark triplet trion. 
The triplet PL intensity is contributed by the bright and dark triplet trions 
\cite{Yusa, Sanvitto}. 
The dark triplet state can recombine 
through a scattering process that changes its total angular momentum.
Indeed, the increase in triplet PL intensity 
due to a random potential induced by remote ionized donors
has been observed \cite{Hayakawa}. 
Therefore, we claim that the similar scenario occurs 
owing to the potential fluctuation induced by polarized nuclear spins.  

In conclusion, 
we demonstrated the simultaneous measurement of resistively and optically detected NMR at $\nu=2/3$.
The simultaneously measured NMR spectra qualitatively showed the same features, 
which are broad and sharp resonances 
respectively associated with spin polarized and unpolarized regions. 
From the unexpected optical NMR spectra, we interpreted 
the optically detected signal 
as a consequence of trion scattering 
induced by polarized nuclear spins. 
Thus, both detection methods probe almost the same local phenomena. 
This means that optical accessible phenomena do not occur in the spatially limited region. 
Even though our identical observations of the NMR spectra with different methods are not surprising naively, 
considering how and where to detect the phenomena by each method provides 
a new insight into the nuclear spin phenomena at $\nu=2/3$, 
which can lead to a future application to quantum information technology.  
We believe that simultaneous measurement by different methods 
and the consideration of their detection details are important 
and give a way to understand physics behind controversial disagreements 
between different detection methods. 

\begin{acknowledgements}
We greatly appreciate K. Muraki for providing the high-quality wafers. 
We also thank T. Yuge, T. Tomimatsu, G. Yusa, and J. Hayakawa for their fruitful discussions. 
Y. H. acknowledges financial support from KAKENHI Grant No. 15H05867.
\end{acknowledgements}


\begin{thebibliography}{99}
\bibitem{Notelocal}
The NMR measures locally at the site of nuclei 
and the collection of this local information forms the NMR spectrum. 
We use the term ``local" in this meaning.
\bibitem{SN}
The maximum change in triplet (singlet) PL intensity is $\sim$2\% ($\sim$0.5\%) of total intensity.
\bibitem{comment1}
There seems to be somewhat different resonances among (a)--(c). 
Although we cannot properly discuss the differences due to the signal fluctuations, 
they may be caused by changes in PL intensity 
due to the sizes of spin polarized and unpolarized regions.
\end{thebibliography}

\end{document}